\documentclass[a4paper,11pt]{article}
\pdfoutput=1 

\usepackage{jcappub} 
\usepackage{hyperref}
\usepackage[T1]{fontenc} 

\newcommand{\E}{\mathcal{E}}

\title{\boldmath Semiclassical geons as solitonic black hole remnants}

\author[a]{Francisco S.N.~Lobo}
\author[b]{Gonzalo J. Olmo}
\author[b,c]{D. Rubiera-Garcia}

\affiliation[a]{Centro de Astronomia e Astrof\'{\i}sica da
Universidade de Lisboa, Campo Grande, Ed. C8 1749-016 Lisboa,
Portugal}
\affiliation[b]{Departamento de F\'{i}sica Te\'{o}rica and IFIC,
Centro Mixto Universidad de Valencia - CSIC. Universidad de
Valencia, Burjassot-46100, Valencia, Spain}
\affiliation[c]{Departamento de F\'isica, Universidade Federal da
Para\'\i ba, 58051-900 Jo\~ao Pessoa, Para\'\i ba, Brazil}

\emailAdd{flobo@cii.fc.ul.pt}
\emailAdd{gonzalo.olmo@csic.es}
\emailAdd{rubieradiego@gmail.com}

\abstract{ We find that the end state of black hole evaporation could be represented by non-singular and without event horizon stable solitonic remnants with masses of the order the Planck scale and up to $\sim 16$ units of charge. Though these objects are locally indistinguishable from spherically symmetric, massive electric (or magnetic) charges, they turn out to  be sourceless geons containing a wormhole generated by the electromagnetic field.  
Our results are obtained by interpreting semiclassical corrections to Einstein's theory in the first-order (Palatini) formalism, which yields second-order equations and avoids the instabilities of the usual (metric) formulation of quadratic gravity. We also discuss the potential relevance of these solutions for primordial black holes and the dark matter problem.}

\begin{document}
\maketitle
\flushbottom

\section{Introduction}
\label{sec:intro}
The idea that gravitationally collapsed objects of very low mass could have been formed by large amplitude density perturbations in the very early universe was proposed by S. Hawking more than 40 years ago \cite{H1}. Conservative estimations, based entirely on the use of classical general relativity (GR), suggested that primordial black holes (PBHs) should exist with masses from $10^{-5}$g upwards. Some years later, Hawking also found that quantum instabilities inherent to the existence of the event horizon render such PBHs quantum mechanically unstable \cite{H2}. This implies that only PBHs with masses above $10^{15}$g could have survived until today \cite{reviewPBHs}. Determination of their current abundance could shed useful light on the primordial power spectrum at very small length scales \cite{PowerSpectrumPBHs}, on the fraction of dark matter that could be attributed to these objects \cite{Kesden:2011ij}, and on their effect on big bang nucleosynthesis \cite{Carr:2009jm}.

Hawking's result on the evaporation of black holes was derived assuming the propagation of quantum fields in the classical background described by GR. In this sense, we note that renormalizability of the matter fields in curved space-times requires a high-energy completion of the Einstein-Hilbert gravitational Lagrangian that involves quadratic curvature terms \cite{parker-toms,birrell-davies}. These contributions generally involve higher-order derivatives of the metric which, besides making it more difficult to find exact solutions, at the perturbative level imply the existence of ghostlike particles that violate unitarity, break causality, and generate inadmissible instabilities \cite{cembranos}.  However, it has been shown that quadratic curvature terms interpreted in a first-order (or Palatini) formalism, which has interesting connections with non-perturbative approaches to quantum gravity  \cite{O&S2009},  yield second-order field equations that in vacuum boil down to those of GR and, consequently, are ghost-free. This property is known as the universality of Einstein's equations in the Palatini approach \cite{Ferraris:1992dx,Borowiec:1996kg}. Therefore, the Palatini formulation avoids fundamental problems present in the more standard (metric) formulation of the semiclassical theory. Given our current understanding of quantum gravity and the lack of evidence supporting that the space-time structure should be Riemannian or otherwise at arbitrarily short distances, we find it worth exploring the effects that assuming metric and  connection as independent entities could have on the predictions of quadratic gravity. Working in the Palatini framework \cite{Palatini-review}, therefore, here we find exact analytical solutions that are in complete quantitative agreement with Hawking's estimates \cite{H1} and, in addition, support the view that the end state of black hole evaporation needs not be pure radiation, that stable remnants with a mass of order the Planck mass, $m_P\sim 2.18\times 10^{-5}$g, may arise. In fact, we find a family of geon-like solitonic objects \cite{Wheeler:1955zz}, i.e., non-singular entities entirely supported by the gravitational and electromagnetic fields, with a discrete mass spectrum given by
\begin{equation}\label{eq:spectrum}
M\approx 1.23605 \left(\frac{N_q}{N_q^c}\right)^{3/2} m_P \ ,
\end{equation}
where $N_q>0$ represents the number of (electric) charges that a local observer would measure and $N_q^c\equiv\sqrt{2/\alpha_{em}}\approx 16.55$,  where $\alpha_{em}$ is the electromagnetic fine structure constant. If $N_q\le N_q^c$, these objects are not hidden behind an event horizon, which makes them stable against Hawking decay.

\section{Theory and solutions}

To proceed, we will first obtain solutions of the theory assuming the existence of a spherically symmetric electric (or magnetic) field. Then we will show that a wormhole structure arises naturally and allows to consistently interpret these solutions as geons. Our theory is described by the following action:
\begin{eqnarray}\label{eq:action}
S[g,\Gamma,\psi_m]&=&\frac{1}{2\kappa^2}\int d^4x \sqrt{-g}\left[R+l_P^2(a R^2+R_{\mu\nu}R^{\mu\nu})\right]
  \nonumber \\
&&-\frac{1}{16\pi}\int d^4x \sqrt{-g}F_{\mu\nu}F^{\mu\nu} \ ,
\end{eqnarray}
where $\kappa^2\equiv 8\pi G/c^4$, $l_P^2\equiv\hbar G/c^3$ represents the Planck length squared, $a$ is a free parameter,  $F_{\mu\nu}$  is the electromagnetic field strength, $g_{\mu\nu}$ is the space-time metric, $R=g^{\mu\nu}R_{\mu\nu}$, $R_{\mu\nu}={R^\rho}_{\mu\rho\nu}=R_{\nu\mu}$, and
${R^\alpha}_{\beta\mu\nu}=\partial_{\mu}
\Gamma^{\alpha}_{\nu\beta}-\partial_{\nu}
\Gamma^{\alpha}_{\mu\beta}+\Gamma^{\alpha}_{\mu\lambda}\Gamma^{\lambda}_{\nu\beta}-\Gamma^{\alpha}_{\nu\lambda}\Gamma^{\lambda}_{\mu\beta} $. The connection $\Gamma^{\alpha}_{\mu\nu}$ is {\it a priori} independent of the metric (Palatini formalism) and must be determined by the field equations  \cite{OSAT09}
\begin{eqnarray}
f_R R_{\mu\nu}-\frac{f}{2}g_{\mu\nu}+2l_P^2 R_{\mu\alpha}{R^\alpha}_\nu &=& \kappa^2 T_{\mu\nu}\label{eq:met-varX}\\
\nabla_{\beta}\left[\sqrt{-g}\left(f_R g^{\mu\nu}+2l_P^2 R^{\mu\nu}\right)\right]&=&0  \ ,
 \label{eq:con-varX}
\end{eqnarray}
where $f=R+l_P^2(a R^2+R_{\mu\nu}R^{\mu\nu})$, $f_R\equiv \partial_R f$, and $T_{\mu\nu}=\frac{1}{4\pi}(F_{\mu\alpha}{F_{\nu}}^{\alpha}-\frac{1}{4}F_{\alpha\beta}F^{\alpha\beta} g_{\mu\nu})$.
The second of these equations follows from variation of the  action with respect to the connection and can be formally solved by means of algebraic manipulations \cite{OSAT09}. The result implies that $\Gamma^{\alpha}_{\mu\nu}$ can be written as the Levi-Civita connection of an auxiliary metric $h_{\mu\nu}$, which is related with $g_{\mu\nu}$ by
\begin{equation} \label{eq:h-g}
h^{\mu\nu}=\frac{g^{\mu\alpha}{\Sigma_{\alpha}}^\nu}{\sqrt{\det\Sigma}} \ , \quad
h_{\mu\nu}=\left(\sqrt{\det\Sigma}\right){\Sigma_{\mu}}^{\alpha}g_{\alpha\nu} \ .
\end{equation}
When $T_{\mu\nu}$  represents a monopolar Maxwell field,  ${\Sigma_{\mu}}^{\nu}$ takes the form (using matrix notation)
\begin{equation}
{\Sigma_{\mu}}^{\nu}=\begin{pmatrix}
\sigma_-\hat{I}& \hat{0} \\
\hat{0} & \sigma_+\hat{I}
\end{pmatrix} \ ,
\end{equation}
where $\hat{I}$ and $\hat{0}$ represent the $2\times2$ identity and zero matrices, respectively, and
\begin{equation}
\sigma_{\pm}=1\pm \frac{r_q^2l_P^2}{ r^4}  ,
\end{equation}
where $r_q^2\equiv {\kappa}^2q^2/4\pi$, $q^2$ is the total charge squared, and we have used that for the electromagnetic field $R=0$ and $R_{\mu\nu}R^{\mu\nu}=r_q^4/r^8$,
which follow from the field equations and coincide with their values in GR. In terms of $h_{\mu\nu}$, Eq.(\ref{eq:met-varX}) boils down to \cite{OR12c}
\begin{equation}
{R_\mu}^\nu(h)=\frac{r_q^2}{2r^4}\begin{pmatrix}
-\frac{1}{\sigma_+} \hat{I}& \hat{0} \\
\hat{0} & \frac{1}{\sigma_-} \hat{I}
\end{pmatrix}  \label{eq:Ricci-h4} \ .
\end{equation}
These equations exactly recover GR in the limit $\hbar\to 0$ (or $l_P\to 0$).   Assuming a spherically  symmetric line element for $h_{\mu\nu}$, the solutions to Eq.(\ref{eq:Ricci-h4}) can be readily found. Transforming this solution back to $g_{\mu\nu}$ using Eq.(\ref{eq:h-g}), one finds that  $ds^2=g_{tt}c^2dt^2+g_{rr}dr^2+r^2d\Omega^2$ is given by
\begin{equation}\label{eq:g}
g_{tt}=-\frac{A(z)}{\sigma_+} \ , \ g_{rr}=\frac{\sigma_+}{\sigma_-A(z)}  \ , \ A(z)=1-\frac{\left[1+\delta_1 G(z)\right]}{\delta_2 z \sigma_-^{1/2}} \ ,
\end{equation}
where $z\equiv r/r_c$, $r_c\equiv \sqrt{r_q l_P}$,  and we have defined
\begin{equation}\label{eq:d1d2}
\delta_1=\frac{1}{2r_S}\sqrt{\frac{r_q^3}{l_P}} \ , \
\delta_2= \frac{\sqrt{r_q l_P} }{r_S} \ .
\end{equation}
Here $r_S\equiv 2GM/c^2$ represents the Schwarzschild radius of the zero charge solution. The function $G(z)$ satisfies
\begin{equation} \label{eq:Gz}
\frac{dG}{dz}=\frac{z^4+1}{z^4\sqrt{z^4-1}} \ .
\end{equation}
Note that expanding Eqs.(\ref{eq:Gz}) and (\ref{eq:g}) far from the center ($z\gg 1$, or $r\gg l_P$), we recover the expected GR limit:
\begin{equation} \label{eq:gtt-far}
g_{tt} = -\left(1-\frac{r_S}{r}+\frac{r_q^2}{2r^2}\right) +\frac{r_q^2 l_P^2}{r^4} +\ldots \ .
\end{equation}
The curvature scalars $R(g)$, $R_{\mu\nu}(g)R^{\mu\nu}(g)$, and ${R^\alpha}_{\beta\mu\nu}(g) {R_\alpha}^{\beta\mu\nu}(g)$ also recover the GR values with corrections $\sim r_q^4l_P^4/r^{10}$, $\sim r_q^4 l_P^2/r^{10}$, and $\sim r_S r_q^2 l_P^2/r^9$, respectively. This puts forward that a few $l_P$ units away from the center, the geometry is virtually indistinguishable from the usual Reissner-Nordstr\"om solution of GR. However, as $z\to 1$ the geometry undergoes important non-perturbative changes, as we will see below. An exact solution of (\ref{eq:Gz}) in terms of infinite power series expansions appears in \cite{OR12c}.\\

\section{Wormhole extension}

The line element  of the metric (\ref{eq:g}) can also be expressed as
\begin{equation}\label{eq:EF}
ds^2=g_{tt}dv^2+ 2dv dr^*+r^2(r^*)d\Omega^2 \ ,
\end{equation}
where $r=r(r^*)$ is such that $(dr^*/dr)^2=-{g_{tt}g_{rr}}=1/\sigma_-$, $v=ct+ x(r)$, and $(dx/dr)^2=-g_{rr}/g_{tt}$.  The line element (\ref{eq:EF})  puts forward that the geometry is fully characterized by the functions $g_{tt}$ and $r(r^*)$. The relation between $z$ and $z^*\equiv r^*/r_c$ can be found by direct integration and is given by $z^*(z)= {_{2}F}_1\left[-\frac{1}{4},\frac{1}{2},\frac{3}{4},\frac{1}{z^4}\right] z$, where ${_{2}F}_1 $ is a hypergeometric function. For $z\gg 1$, we have $z^*\approx z$, whereas for $z\to 1$ we find $z^*\approx z^*_c +\sqrt{z-1}+\frac{5}{12} (z-1)^{3/2}+\ldots$, where $z^*_c\approx 0.59907$.  The relation between $z$ and $z^*$ is monotonic and invertible in the region $z\ge 1$. From (\ref{eq:Gz}) we also see that $G(z)$ is only defined for $z\ge 1$. Thus, the radial coordinate $r$ has the range $r_c \leq r < +\infty $, which is reminiscent of a wormhole geometry, where $r=r_c$ (or $z=1$) would correspond to the presence of a wormhole throat. To better understand the geometry, we expand  around $z\approx 1$ obtaining
\begin{equation}
g_{tt}= \frac{\left(1-\delta _1/\delta_1^*\right)}{4\delta _2 \sqrt{z-1}}-\frac{1}{2}\left(1-\frac{\delta _1}{ \delta _2}\right)+O(\sqrt{z-1}) \label{eq:gtt_series} \ ,
\end{equation}
where $\delta_1^* \approx 0.5720$ is a constant needed to match the series expansions of $G(z)$ in the $z\to 1$ and $z\gg 1$ regions. Though the expression (\ref{eq:gtt_series}) is in general divergent as $z\to 1$, for the particular choice $\delta _1=\delta _1^*$ the expansion is regular and yields a smooth geometry.  This is also supported by the fact that for $\delta_1=\delta_1^*$ the Kretschmann scalar of $g_{\mu\nu}$ at $z=1$ is finite:
\begin{equation}
{R^\alpha}_{\beta \mu\nu}(g){R_\alpha}^{\beta \mu\nu}(g)=\frac{1}{(r_q l_P)^2}\left(16-\frac{32 }{3 }\frac{ r_q}{ l_P}+\frac{22 }{9 }\frac{ r_q^2}{ l_P^2}\right)  \label{eq:Kret} \ .
\end{equation}
$R(g)$ and $R_{\mu\nu}(g)R^{\mu\nu}(g)$ are also finite in this case.

\begin{figure}[tbp]
\centering 
\includegraphics[width=0.5\textwidth]{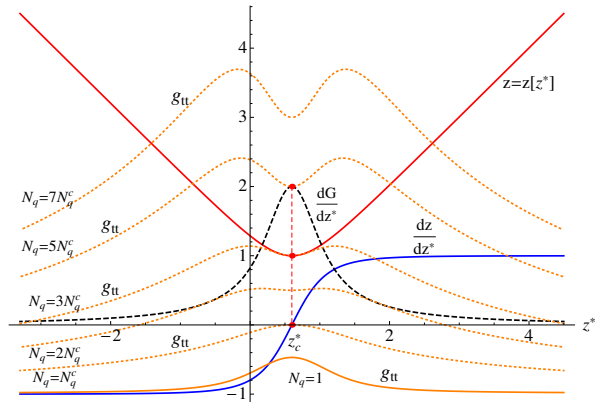}
\caption{The minimum of $z(z^*)$ occurs at $z^*_c\approx 0.599$, where $dz/dz^*$ vanishes, $dG/dz^*$ reaches its maximum, and $g_{tt}$ has an extremum. We have plotted $g_{tt}$ with  $\delta_1=\delta_1^*$ for $N_q=1$ (solid orange) and $N_q= N_q^c, 2N_q^c, 3N_q^c, 5N_q^c, 7N_q^c$ (dotted orange). Note that for $N_q=1$ we have $g_{tt}\approx -1$ for $z\gtrsim 2$. 
\label{fig:zzstar}}
\end{figure}

It must be noted that though in the construction of $z^*(z)$ we have implicitly assumed that $dz^*/dz=   1/\sigma_-^{1/2}$, for $\delta_1=\delta_1^*$, a solution with $dz^*/dz=  - 1/\sigma_-^{1/2}$ is also possible. Moreover, since for $\delta_1=\delta_1^*$ the geometry at $z=1$ is smooth [see Eqs.(\ref{eq:gtt_series}) and (\ref{eq:Kret})],  the divergence of $dz^*/dz$
at this point simply states that the function $z(z^*)$ has reached a minimum at $z^*_c$ (see Fig. \ref{fig:zzstar}). For values of $z^*<z^*_c$, the branch with  $dz^*/dz<0$ describes a new region across the throat of the wormhole, in which the area of the $2-$spheres grows as $z^*\to -\infty$.  The relation between $z$ and $z^*$ is thus given by (see Fig. \ref{fig:zzstar})
\begin{equation}
z^*=\left\{\begin{array}{lr}
 {_{2}F}_1\left[-\frac{1}{4},\frac{1}{2},\frac{3}{4},\frac{1}{z^4}\right] z & \text{ if } z^*\ge z^*_c \\
2z^*_c-{_{2}F}_1\left[-\frac{1}{4},\frac{1}{2},\frac{3}{4},\frac{1}{z^4}\right] z & \text{ if } z^*\le z^*_c
\end{array}\right.
\end{equation}
The coordinate $r^*=r_c z^*$ is thus defined from $-\infty$ to $+\infty$, and the area $A=4\pi r^2(r^*)$ of the $2-$spheres undergoes a bounce after reaching a minimum at $z^*=z^*_c$, which manifests a genuine wormhole structure connecting two regions through a (spherical) tunnel of radius $r_c$. Unlike other known wormhole solutions that require exotic matter sources to generate a pre-designed geometry \cite{WHmodgrav}, our solution comes out naturally from the field equations and is entirely supported by an electric field.

\section{Geon structure and solitonic interpretation}

The smooth wormhole structure of the solutions with $\delta_1=\delta_1^*$ implies that the lines of force of the electric field enter through one of the wormhole mouths and exit through the other creating the illusion of a negatively charged object on one side and a positively charged object on the other. The locally measured electric charge can be obtained by computing the flux $\Phi\equiv \int_S *F=4\pi q$ through any $2-$surface $S$ enclosing the wormhole throat, where $*F$ represents the 2-form dual to Faraday's tensor (note, in particular, that if $S$ is taken as a $2-$sphere, the respective normal vectors to the $2-$spheres pointing in the direction of increasing $r$ on both sides of the wormhole differ by a sign, which explains the different sign of the charge measured locally).
This shows that  no {\it real} sources generate the field, which is fully consistent with the sourceless gravitational-electromagnetic equations of our theory (\ref{eq:action}) and with Wheeler's definition of geons \cite{Wheeler:1955zz}. 
It is worth noting that the electric flux per surface unit at $z=1$, $\Phi/4\pi r_c^2$, which represents the density of lines of force crossing the wormhole throat, turns out to be a universal quantity, $\Phi/4\pi r_c^2 = q/ r_c^2=\sqrt{c^7/(\hbar G)^2}/\sqrt{2}$, that only depends on $\hbar$, $c$, and $G$. The fact that this quantity is independent of the charge and mass of the particular solution considered supports the view that the geon structure is also present when $\delta_1\neq \delta_1^*$, i.e., that $z=1$ always defines a wormhole throat traversed by a sourceless electric flux. This puts forward that the space-time geometry can be extended to the $z^*<z^*_c$ region even for solutions with $\delta_1\neq \delta_1^*$, for which curvature scalars diverge at $z=1$ (the divergence goes as  $\sim 1/(z-1)^3$ and is much weaker than in GR, $\sim 1/r^8$).

Having extended the geometry to the whole range of $r^*$ (the real axis), one finds that the addition of the electromagnetic energy stored in the field, $\E_{e.m.}= -\int_{-\infty}^{\infty}dr^* r^2d\Omega \frac{1}{16\pi} F_{\mu\nu}F^{\mu\nu}$, with the gravitational binding energy (as given by the evaluation of the gravitational action on the solution), turns out to be $\E_{Tot}=\frac{q^2}{r_c \delta_1^*}=2\frac{\delta_1}{\delta_1^*}M c^2 $. Remarkably, taking into account the existence of the two sides of the wormhole,  this result indicates that when $\delta_1=\delta_1^*$ the gravitational mass of the system as locally measured on one side of the wormhole is entirely due to the energy generated by the electric field on that side of the wormhole, $Mc^2=\E_{Tot}/2$, which naturally allows to interpret such solutions as geonic solitons.

\section{Stability and quantum properties}

Though the solitonic solutions just found are classically stable for topological reasons \cite{Misner-Wheeler1957}, it is also true that quantum instabilities due to the existence of an event horizon may force their decay into states with $\delta_1\neq \delta_1^*$. In this respect, a numerical search of the horizons using the exact solution of (\ref{eq:Gz}) when $\delta_1=\delta_1^*$ shows that the sign of the term $\left(1-{\delta_1^*}/{\delta _2 }\right)$ in (\ref{eq:gtt_series}) determines whether an event horizon exists or not  \cite{OR12c}. Since  $\delta_1^*/\delta_2=r_q/2l_P$, it follows that the event horizon is absent if $r_q<2l_P$. This inequality can be written as a constraint on the charge of the system. In fact, expressing the charge as $q=N_q e$, where $e$ is the electron charge and $N_q$ the number of charges,  one finds that $r_q=2l_P N_q /N_q^c$, where $N_q^c\approx 16.55$ was defined below Eq.(\ref{eq:spectrum}), which leads to $\delta_1^*/\delta_2\equiv N_q/N_q^c$. Therefore, for $N_q>N_q^c$ an event horizon exists (and its location is almost coincident with the prediction of GR  for $N_q\gtrsim 30$ \cite{OR12c}). However, objects with $N_q<N_q^c$  have no event horizon, which guarantees their stability against Hawking decay. Moreover, the regularity condition $\delta_1=\delta_1^*$, which according to (\ref{eq:d1d2}) establishes a constraint between the mass and the charge of the object (and also the identification of gravitational mass with the soliton energy), can be rewritten to yield Eq.(\ref{eq:spectrum}), which sets a minimum mass of $M_{q=1}\approx m_P/55$.

The fact that solitonic states stable against Hawking decay exist in the lowest band of the mass and charge spectrum of our theory strongly supports the view that the end state of black hole evaporation might be represented by these objects. The condition $\delta_1=\delta_1^*$ could thus be seen as a quantum rule that selects a discrete set among the classically allowed solutions, similarly as stable orbits arise in Bohr's atomic model.

\section{Concluding remarks}

The potential existence of stable massive particles has been thoroughly investigated in the last years in connection with the dark matter problem \cite{DM}. The results presented here offer a gravitational alternative to this issue in the form of horizonless geonic solitons stable against Hawking decay [see Eq. (\ref{eq:spectrum})]. Though the innermost structure of these objects is non-trivial, for scales larger than a few times $l_P$ they are virtually indistinguishable from the usual charged Reissner-Nordstr\"{o}m black holes of GR [see Eq.(\ref{eq:gtt-far})]. In this sense, it should be noted that our results are fully compatible with Hawking's original estimates, who found that primordial black holes with masses larger than $\sim 10^{-5}$g and up to $\sim 30$ charges could be formed through classical collapse \cite{H1}. Therefore, following Hawking's prediction, one could expect that a considerable amount of such stable objects could have been produced in the early universe \cite{PBH}. From our solutions, however, it might be argued that the electric field at the wormhole throat is sufficiently intense as to induce the creation of pairs out of the vacuum, whose existence would alter our results (an issue that also applies to GR). Though such quantum polarization effects have been neglected in this work (like in the standard analysis in GR), it will be shown elsewhere \cite{O&R13} that the qualitative picture provided by our model is  robust against quantum corrections coming from the matter sector. In fact, assuming that the matter quantum corrections end up generating nonlinear contributions in the electromagnetic sector, one can explicitly show that the wormhole structures found here persist. Moreover, the resulting mass spectrum can be largely reduced by many orders of magnitude due to the nonlinear matter corrections. As a result, stable solitonic structures such as those presented here with massess within the observational constraints established for charged massive particles (CHAMPs) \cite{Chuzhoy:2008zy} could have been naturally produced in the early universe.

As Hawking pointed out, these objects could become neutral and non-relativistic by capturing free charges to form ultra-heavy atoms (see also \cite{Chuzhoy:2008zy} for more details on the expected phenomenology of these objects).  On the other hand, since our theory admits both electric and magnetic monopolar solutions, the combination of pairs with opposite charges into bound states may also provide another source of massive neutral particles, which we tentatively denote {\it geonium}. In this sense, it has been argued \cite{Vento} that the binding energy of two magnetic monopoles ({\it monopolium}) could significantly reduce the energy threshold required to generate the pair, which could facilitate the production and detection of {\it geonium} in particle accelerators \cite{Moedal}. These objects are also likely to arise in the last stages of black hole evaporation, when large amounts of energy are expected to be radiated away through the emission of all kinds of particles. In summary, due to their heavy mass and stability against Hawking decay, our solutions represent natural candidates for PBHs and black hole remnants \cite{Adler:2001vs} and, consequently, for dark matter. Their existence naturally implies a maximum temperature for the black hole evaporation process, which could justify the lack of observational evidence for black hole explosions \cite{EGRET}. 

To conclude, we note that our model combines ideas coming from the quantization of matter fields in curved backgrounds \cite{parker-toms,birrell-davies} with others that have allowed important progress in the canonical non-perturbative quantization of GR \cite{ashtekar}, namely, that when metric and connection are regarded as independent geometrical entities unexpected aspects of the microstructure of the space-time may arise. Though  the semiclassical aspects of Palatini theories are not yet well understood, the analytical tractability of the model (\ref{eq:action}) and the potential implications that the results obtained within this framework may have for the understanding of dark matter, black holes, and the very structure of space-time and elementary particles justify its interest and motivate further research in this direction.

\acknowledgments

 F.S.N.L. acknowledges financial support of the FCT through grants CERN/FP/123615/2011 and CERN/FP/123618/2011. G.J.O. is supported by the Spanish grant FIS2011-29813-C02-02 and the JAE-doc program of the Spanish Research Council (CSIC). D.R.-G. is supported by CNPq through grant 561069/2010-7, and thanks the hospitality and partial support of the theoretical physics group at the University of Valencia.

\end{document}